\documentclass[10pt]{article}

\usepackage{aas2pp4}
\usepackage{amssym}
\usepackage{flushrt}
\usepackage[subnum]{cases}                                 

\newcommand{\rd}{{\rm d}}
\lefthead{Querella}
\righthead{CFS kinematic cosmology}
 
\begin{document}

\title{Kinematic cosmology in conformally flat spacetime}

\author{Laurent Querella}

\affil{Boursier {\sc fria}%
          \footnote{Fonds pour la formation \`{a} la recherche dans 
                    l'industrie et dans l'agriculture}, \\
       Groupe de cosmologie, th\'{e}orie des champs et particules 
       fondamentales, \\
       Institut d'astrophysique et g\'{e}ophysique de l'universit\'{e} de 
       Li\`{e}ge \\
       Avenue de Cointe 5, B--4000 Li\`{e}ge, Belgique. \\
       E--mail~: L.Querella@ulg.ac.be}

\authoraddr{Institut d'astrophysique et g\'{e}ophysique, universit\'{e} de 
            Li\`{e}ge, avenue de Cointe 5, B--4000 Li\`{e}ge, Belgique} 

\begin{abstract}
In a recent series of papers Endean examines the properties of spatially
homogeneous and isotropic (FLRW) cosmological models filled with dust in the 
``conformally flat spacetime presentation of cosmology'' (CFS cosmology). 
This author claims it is possible to resolve a certain number of the 
difficulties the standard model exhibits when confronted to observations, if 
the theoretical predictions are obtained in the special framework of CFS 
cosmology. As a by-product of his analysis Endean claims that no initial 
(big-bang) nor final (big-crunch) singularities occur in the closed FLRW 
model. In this paper we show up the fallacious arguments leading to Endean's 
conclusions and we consistently reject his CFS cosmology.
\end{abstract}


\section{Introduction}

Kinematical \, aspects \, of \, Friedmann\---Lema\^\i tre\---Robertson\---%
Walker (FLRW) models have been examined by \cite{infe} (1945). These authors 
determined exhaustively all the possible forms of the metrics written 
explicitly in conformally flat form and classified all the different types of 
motion of free particles and light rays in these various universes. In their 
study they willfully ignored the actual dynamics of the cosmological models. 
This very problem was tackled by \cite{taub} (1967) who solved explicitly 
Einstein's equations for the FLRW conformally flat form metrics and for 
various types of equation of state. \\

The conformally flat spacetime presentation of cosmology (CFS cosmology) has 
been developed in a recent series of papers. It differs from Tauber's approach 
since it takes root in \cite{infe}'s kinematic cosmology and does not analyze 
the dynamical behaviour of the conformal factor. \cite{ende1} (1994) examines 
the consequences of expressing redshift-distance relations and the timing test 
in terms of the CFS coordinates for the open FLRW model filled with dust. 
\cite{ende2} (1995) extends his previous analysis to the closed FLRW dust 
model taking into account partial determinations of the Hubble constant and 
statistical analysis of redshift-distance data at intermediate range. He 
claims that CFS cosmology resolves the cosmological age and redshift-distance 
difficulties exhibited by these observational data. \cite{ende3} (1997) 
asserts that CFS time provides the true measure of elapsed time in the 
universe and that no spacetime singularities occur in the closed FLRW dust 
model. \\ 
 
In this paper we briefly summarize the key results of kinematic cosmology 
along the lines of \cite{infe} (1945). We point out the interpretational 
difference between this framework and Endean's CFS cosmology. On this basis we 
show up the fallacious arguments underpinning Endean's conclusions, and 
accordingly we reject Endean's spurious cosmology.

\section{Kinematic cosmology}

\subsection{Equivalence of FLRW and conformally flat metrics}

Kinematic cosmology rests on the property that FLRW models are conformally 
flat, i.e. related to some regions of flat spacetimes by means of a conformal 
transformation of the metric, viz. $\widetilde{g}_{ab}=\varphi^2(x) g_{ab}$ 
where $\varphi$ is the conformal factor. Without performing such 
transformation it is always possible to find a transformation of coordinates 
that brings the initial FLRW metrics to manifestly conformally flat forms. One 
starts from the angular form of FLRW metrics in comoving coordinates 
$(\tau, \rho)$, 
\begin{equation}
   \rd s^2 = - \rd \tau^2 
             + R^2(\tau) \left[ 
                            \rd \rho^2 + \chi^2(\rho) \rd \omega^2 
                         \right] , 
\label{flrwcom}
\end{equation}
where $R$ denotes the scale factor of the universe, $\tau$ is the proper 
cosmic time, $\rho$ is the radial comoving coordinate, $\omega$ stands for the 
spherical solid angle, and the functions $\chi(\rho)$ read explicitly for the 
closed ($k=+1$), flat ($k=0$) and open ($k=-1$) models respectively as
$\sin \rho$, $\rho$ and $ \sinh \rho$.
Upon introducing the `development angle' $\eta$ as a new time coordinate, 
$\rd \tau = \pm R(\eta) \rd \eta$, and null coordinates $(\xi, \lambda)$ by
$\xi = \case12 ( \eta + \rho )$, and $\lambda = \case12 ( \eta - \rho )$, the 
metric (\ref{flrwcom}) takes the form
\begin{equation}
   \rd s^2 = R^2 \left(\xi + \lambda \right)
             \left[ - 4 \rd \xi \rd \lambda 
                    + \chi^2 \left( \xi - \lambda \right) \rd \omega^2 
             \right]. 
\label{flrwint}
\end{equation}
The transformation that turns the metric (\ref{flrwint}) into a conformally 
flat metric is defined by $X = g(\xi) = t + r$ and $Y = g(\lambda) = t - r,$ 
where $(t,r)$ are the CFS coordinates, and the function $g(w)$ is given 
explicitly for the closed, flat and open models respectively by $\tan w$, $w$ 
and $\tanh w$ (\cite{ligh} 1975). With this transformation the metric 
(\ref{flrwint}) becomes
\begin{equation}
   \rd s^2 = \gamma (r,t) 
             \left[ - \rd t^2 + \rd r^2 + r^2 \rd \omega^2 \right], 
\label{flrwcfs}
\end{equation}
where the conformal factor satisfies the identity
\begin{equation}
   \gamma (r,t) = \frac{4 R^2 \left[ g^{-1} (t + r) + g^{-1} (t - r) \right]}   
                       { g^{\prime} (t + r) g^{\prime} (t - r)},
\label{cfacident}
\end{equation}
$g^{\prime}$ and $g^{-1}$ denoting respectively the derivative of $g(w)$ and 
the inverse function to $g(w)$. The relationship between the comoving 
coordinates $(\eta, \rho)$ and the CFS coordinates is readily obtained from 
the above definitions. It reads
\begin{equation}
   t \pm r = g \left( \frac12 (\eta \pm \rho) \right).
\label{comcfs}
\end{equation}
At this stage it is important to note that if one takes up the CFS metric
(\ref{flrwcfs}) instead of the conventional metric (\ref{flrwcom}) one shifts
the study of the space structure, i.e. the knowledge of the scale factor, 
which characterizes the dynamics of the model under consideration to the 
analysis of the types of motion of fundamental particles (kinematics). In 
total agreement with \cite{infe} (1945) we stress here that these types of 
motions have no straightforward physical interpretation. If one's aim is to 
study the dynamics in CFS coordinates one has to follow \cite{taub}'s (1967) 
approach, i.e. solve Einstein's equations for the conformal factor 
$\gamma(r,t)$. The would-be advantage of choosing the CFS metric 
(\ref{flrwcfs}) lies in the fact that light rays propagate on straight lines 
with constant velocity in CFS coordinates (because conformal transformations 
preserve the light cone structure --- of Minkowski spacetime in this case). In 
the line of \cite{infe} (1947) one can set a good illustration of this 
property by deriving a general formula for the redshift of distant objects in 
CFS coordinates.

\subsection{General formula for the redshift}

Consider an atom, moving with a fundamental particle $P=(r,t)$ of radial
velocity $v(r,t)=\rd r/ \rd t$ and emitting photons with a proper period 
$\rd \tau$ in the direction of an observer at $(0,t_0)$, where $t_0$ is the 
present CFS time. The corresponding null geodesics of the CFS metric 
(\ref{flrwcfs}) are straight lines, $t_0=t+r$. The proper emitted wavelength 
is $\lambda_{\rm e} = \rd \tau = \gamma^{\slantfrac12} \sqrt{1-v^2} \, \rd t$,
and the proper observed wavelength is 
$\lambda_0 = \rd \tau_0 = \gamma^{\slantfrac12}(0,t_0) (1+v) \rd t$. 
Therefore the redshift is given by
\begin{equation}
   1 + z = \frac{\lambda_0}{\lambda_{\rm e}}
         = \left(\frac{\gamma_0}{\gamma}\right)^{\slantfrac12}
           \left(\frac{1+v}{1-v}\right)^{\slantfrac12}.
\label{redcfs}
\end{equation}          
One clearly sees that in CFS coordinates there are two contributions to the 
redshift, namely the gravitational and Doppler effects. Note that equation
(\ref{redcfs}) could be obtained by performing the transformation 
(\ref{comcfs}) on the standard formula
\begin{equation}
   1 + z = \frac{R(\eta_0)}{R(\eta)}.
\label{redcom}
\end{equation}          
At this stage we arrive at the very key-stone of Endean's work since this
author looks at the consequences of expressing the redshift formula 
(\ref{redcom}) in terms of the CFS coordinates. There are no a priori 
objections against such a procedure provided one bears in mind the unphysical 
nature of CFS coordinates. However, Endean overrules the fundamental postulate 
of general relativity, namely the general covariance of the physical laws, by 
ascribing an absolute physical meaning to CFS coordinates. It should be clear
that this infringment is simply not acceptable for there is no reason
whatsoever to forsake the coordinate invariance of physical laws. For 
instance, \cite{ende3} (1997) claims that CFS time $t$ correctly measures the 
true age of the universe whereas, by definition, the genuine elapsed time is 
the proper cosmic time $\tau$. Moreover, in contrast to \cite{infe} (1945), 
Endean merges purely kinematic concepts into dynamical ones, asserting that 
the radial velocity $v(r,t)$ may be interpreted as the cosmological recession 
velocity. Underpinning his arguments on such mistaken assumptions conveys 
Endean to startling conclusions as for instance the vanishing of spacetime 
singularities (see \S 3.3.). 
 
\section{CFS cosmology}

For the sake of briefness we now restrict ourself to the closed FLRW 
cosmological model with dust in order to show up the fallacies present in 
\cite{ende3} (1997). For our purpose we focus the discussion on the behaviour 
of the scale factor and of the cosmological parameters expressed in CFS 
coordinates. 
 
\subsection{Scale factor and $t$-clock}

The exact solution of Einstein's equations describing a closed FLRW model
filled with dust can be found in any textbook on cosmology and reads, in terms 
of $\eta$, 
\begin{mathletters}
   \begin{eqnarray}
      R(\eta) &=& K (1 - \cos \eta), \\
      \tau    &=& K (\eta - \sin \eta),
   \end{eqnarray}
   \label{sol}   
\end{mathletters}
where $K$ is a constant of integration which could be determined in principle 
from the knowledge of the present values of the Hubble constant $H_0$ and the 
mass density of the universe, 
$ K = \Omega_0 / 2 H_0 \left( \Omega_0 - 1 \right)^{3/2}$
with $\Omega_0$ denoting the present density parameter. On the other hand, the 
time flow in CFS coordinates is not the same as the proper cosmic time flow.
The relationship between a $\tau$-clock and a $t$-clock is indeed given by
\begin{mathletters}
   \begin{eqnarray}
   \rd \tau_0 &=& \gamma^{\slantfrac12}(0, t_0) \rd t_0
               = K \sin^2 \eta_0 \, \rd t_0, \\  
          t_0 &=& \tan \left( \frac{\eta_0}{2} \right). \label{tcl}
   \end{eqnarray}
   \label{tclock}
\end{mathletters}
Turning aside from the interpretation of equations (\ref{tclock}) \cite{ende3} 
(1997) states a new ``principle of identical clock rates'' asserting that the 
flows of CFS and cosmic times are identical, viz. $\rd \tau_0 \equiv \rd t_0$. 
This, a major offense, shows unambiguously that CFS cosmology is fundamentally
unsound. For instance, this ``principle'' has the awkward consequence that $K$ 
is no longer constant but determined as a function of $\eta$, namely 
$K(\eta)=1/\sin^2 \eta$. This is in total contradiction with the exact 
solution (\ref{sol}). In that case the (wrong) scale factor becomes 
$R(\eta)=1/(1+\cos \eta)$ and tends to infinity as $\eta$ tends to $\pi$. On 
account of these misconceptions \cite{ende3} (1997) deduces that the scale 
factor can not reach its maximum value either in a finite CFS time or in a 
finite cosmic time, and that this maximum is infinite. As another consequence 
of this ``principle'' Endean finds that the conformal factor must fulfill the 
condition $\gamma(0,t_0)=1$ for all $t_0$. He then infers from this result a 
property he calls a ``strong Copernician principle'' enabling him to conclude 
that CFS coordinates measure the true distances and elapsed times in the 
universe and that these measures do not depend on the location of the 
observers. The ``strong Copernician principle'' is obviously wrong, for the
conformal factor (\ref{cfacident}) becomes in the closed FLRW model, 
\begin{eqnarray}
   \gamma(r,t) &=& 4 K^2 \left[ 
                            \sqrt{1+X^2} \sqrt{1+Y^2} - (1 - XY) 
                         \right]^2 / \nonumber \\
               & & \quad \left[ 1+X^2 \right]^2 \left[ 1+Y^2 \right]^2,
\label{cfacCFS}
\end{eqnarray}
where $X, Y$ have been defined above and $\gamma(r,t) = 1$ only for $r=0$ and 
$t=1.$

\subsection{Cosmological parameters}

From the exact solution (\ref{sol}) it is straightforward to obtain the Hubble
and deceleration parameters as functions of $\eta$. They read respectively
\begin{mathletters}
   \begin{eqnarray}
      H(\eta) &:=& \frac{1}{R} \frac{\rd R}{\rd \tau} 
                = \frac{\sin \eta}{K (1 - \cos \eta)^2}, 
      \label{hubeta} \\
      q(\eta) &:=& - \frac{1}{R H^2} \frac{\rd^2 R}{\rd \tau^2}
                = \frac12 \left[ 
                             1 + \tan^2 \left( \frac{\eta}{2} \right)   
                          \right].
      \label{qeta}
   \end{eqnarray}
\end{mathletters}
Taking equation (\ref{tcl}) into account their present values in terms of a 
$t$-clock are readily obtained, 
$H_0 = (1 + t_0^2)/(2 K t_0^3),$ and $q_0 = (1 + t_0^2)/2.$   
A usual redshift-distance relation is written down upon expanding the redshift 
formula (\ref{redcom}) around small proper distances $l,$
\begin{equation}
   z \sim H_0 l + \frac12 (1 + q_0) H_0^2 l^2 + {\cal O}(l^3).
\label{redl}
\end{equation}          
If the expansion is performed around small CFS radial coordinates $r$, one
obtains a similar expression, viz.
\begin{equation}
   z \sim \frac2{t_0 \left(1 + t_0^2 \right)} r + 
          \frac{5 t_0^2 + 3}{t_0^2 \left(1 + t_0^2 \right)^2} r^2 + 
          {\cal O}(r^3).
\label{redr}
\end{equation}          
In analogy with equation (\ref{redl}) \cite{ende1} (1994) identifies the 
coefficient in front of the first order term in equation (\ref{redr}) as a new 
Hubble parameter $\bar{H}_0$ which would be the observed Hubble parameter if 
CFS coordinates were physically meaningful. As we know that it is not the 
case, the introduction of $\bar{H}_0$ is completely irrelevant. (Notice also
that the ``principle of identical clock rates'' implies $\bar{H}_0 = H_0.)$
\cite{ende3}'s (1997) subsequent discussion on the timing test in CFS 
coordinates is therefore unnecessary. 

\subsection{Spacetime singularities}

The occurrence of spacetime singularities is a gen\-er\-ic feature of spatially
homogeneous models (\cite{hawk} 1973). All FLRW cosmological models exhibit a
singularity in the past (big-bang) where at $\tau=0$ the mass-energy density
goes to infinity. In addition, the closed model recollapses towards a 
singularity in the future (big-crunch) after a finite proper time. In total
contradiction with these facts, \cite{ende3} (1997) concludes that neither the 
big-bang nor the big-crunch take place in the closed FLRW model if it is 
examined from the point of view of CFS coordinates. With regard to the initial
singularity he grounds his reasoning on the fact that the radial CFS velocity 
$v(r,t)$ (mistakenly considered as the recession velocity), vanishes when 
$t \rightarrow 0.$ We prove now, if necessary, that the initial singularity 
fairly exists in CFS coordinates. For our purpose we consider the curvature 
invariant constructed on the Riemann tensor, viz. $I = R_{klmn} R^{klmn}$. Its 
expansion as the CFS time tends to zero is given at the leading order by 
\begin{equation}
   I \sim \frac{15}{16 K^4} \left(1 + r^2 \right)^{12} t^{-12} + 
          {\cal O} \left( t^{-10} \right), 
\end{equation}
and thus diverges. Note that \cite{taub} (1967) also found a diverging energy 
density. Concerning the final singularity now, it is obvious from equation 
(\ref{tcl}) that if one ascribes a physical meaning to a $t$-clock as in 
\cite{ende3} (1997) then the universe will never be seen to recollapse since 
$t_0 \rightarrow \infty$ as $\eta_0 \rightarrow \pi$, the value $\eta_0=\pi$ 
corresponding to the maximum radius (see eq. [\ref{sol}]). Therefore, Endean's 
claim of a non-recollapsing closed universe is also totally wrong. 

\section{Discussion}

In this paper we have shown up the fallacious assumptions founding Endean's
CFS cosmology. Our conclusions can be summarized as follows. 
First of all, CFS coordinates do not measure true distances and elapsed times 
in the universe. This is an obvious statement on account of the definition of 
CFS coordinates and of the mere inspection of the CFS metric (\ref{flrwcfs}).
(Coordinates are nothing else than labels for events occurring in spacetime,
and it is the metric, eventually, that tells us how to measure distances and
time intervals.) On the one hand Endean's opposite claim with respect to 
elapsed times rests on his totally unjustified ``principle of identical clock 
rates.'' On the other hand his assertion stating that the CFS radial distance 
represents a true measure of distance is partly based on a direct consequence 
of this ``principle,'' and partly justified from a suitable fitting of 
redshift-distance curves (\cite{ende2} 1995). We stress however that 
reproducing curve shapes can never replace a reliable theoretical framework. 
Secondly, the occurrence of singularities in FLRW cosmological models is by no 
means altered upon introducing CFS coordinates --- Physical results are 
coordinate invariant. Endean's opposite conclusion reveals that his 
argumentation rests on the serious misconceptions mentioned above concerning 
the absolute physical meaning ascribed to CFS coordinates which breaks the
general covariance principle. 
On account of these facts we conclude that Endean's CFS cosmology is totally 
unsound and ought to be discarded.


\begin{thebibliography}{7}    

\bibitem[Endean]{ende1}
Endean, G. 1994, Ap. J., 434, 397 
\bibitem[Endean]{ende2}
Endean, G. 1995, Mon. Not. R. Astron. Soc., 277, 627 
\bibitem[Endean]{ende3} 
Endean, G. 1997, Ap. J., 479, 40 
\bibitem[Hawking \& Ellis]{hawk}
Hawking, S. W., \& Ellis G. F. R. 1973, The large scale structure of 
space-time (Cambridge~: Cambridge University Press) 
\bibitem[Infeld \& Schild]{infe}
Infeld, L., \& Schild A. 1945, Phys. Rev., 68, 250 
\bibitem[Lightman et al.]{ligh} 
Lightman, A. P., Press, W. H., Price, R. H., \& Teutolsky, S. A. 1975, Problem
book in relativity and gravitation (Princeton~: Princeton University Press), 
p. 524 
\bibitem[Tauber]{taub}
Tauber, G. E. 1967, J. Math. Phys., 8, 118 

\end{thebibliography}
\end{document}